\documentclass[twocolumn]{IEEEtran} 

\usepackage{amsmath,amssymb,amsfonts}
\usepackage[final]{graphicx}
\usepackage{psfrag}
\usepackage[tight]{subfigure}
\usepackage[numbers,sort&compress]{natbib}

\usepackage{mathrsfs} 
\usepackage{epstopdf}
\usepackage{multirow}
\usepackage{amsthm,mathtools}
\usepackage{float}
\usepackage{booktabs}  
\usepackage{multirow} 
\usepackage{relsize}
\usepackage{bm}
\usepackage[nolist,nohyperlinks]{acronym}
\usepackage{svg}
\usepackage{xparse}
\NewDocumentCommand{\evalat}{sO{\big}mm}{%
  \IfBooleanTF{#1}
   {\mleft. #3 \mright|_{#4}}
   {#3#2|_{#4}}%
}
\newcommand{\norm}[1]{\left\lVert#1\right\rVert}

\begin{acronym}
\acro{PLS}{physical layer security}
\acro{SNR}{signal-to-noise ratio}
\acrodefplural{SNR}[SNRs]{signal-to-noise ratios}
\acro{AWGN}{additive white Gaussian noise}
\acro{CSI}{channel state information}
\acro{PDF}{probability density function}
\acrodefplural{PDF}[PDFs]{probability density functions}
\acro{CDF}{cumulative distribution function}
\acrodefplural{CDF}[CDFs]{cumulative distribution functions}
\acro{LRS}{large reflecting surface}
\acro{BS}{base station}
\acro{RV}{random variable}
\acro{FN}{folded normal}
\acro{TDD}{time-division duplexing}
\acro{LOS}{line-of-sight}
\acro{DL}{downlink}
\acro{MRT}{maximal ratio transmission}
\acro{MRC}{maximal ratio combining}
\acro{MC}{Monte Carlo}
\end{acronym}

\newtheorem{proposition}{Proposition}

\newtheorem{remark}{Remark}

\newcommand{\diag}{\mathop{\mathrm{diag}}}

\usepackage[numbers,sort&compress]{natbib}

\DeclareMathOperator{\sinc}{sinc}

\makeatletter
\def\blfootnote{\xdef\@thefnmark{}\@footnotetext}
\makeatother

\usepackage{float}
\usepackage{stfloats}
\usepackage{textcomp}
\usepackage[ruled,vlined]{algorithm2e}

\begin{document}
\title{\huge{Expectation-Maximization Learning for Wireless Channel Modeling of Reconfigurable Intelligent Surfaces}}
\author{Jos\'e~David~Vega~S\'anchez, Luis~Urquiza-Aguiar, Martha Cecilia Paredes Paredes, and F. Javier L\'opez-Mart\'inez }

\maketitle

\blfootnote{\noindent Manuscript received MONTH xx, YEAR; revised XXX. The review of this paper was coordinated by XXXX. The work of J.~D.~Vega~S\'anchez, L. F. Urquiza-Aguiar, and M. C. Paredes Paredes was funded by the Escuela Polit\'ecnica Nacional (EPN), for the development of the project PIGR-19-06. J.~D.~Vega~S\'anchez received a teaching assistant scholarship for doctoral studies. The work of F.J. Lopez-Martinez was funded by the Spanish Government and the European Fund for Regional Development FEDER (TEC2017-87913-R, PID2020-118139RB-I00) and by Junta de Andalucia (P18-RT-3175).}

\blfootnote{\noindent J.~D.~Vega~S\'anchez, L.~F.~Urquiza-Aguiar and M.~C.~Paredes Paredes are with Departamento de Electr\'onica, Telecomunicaciones y Redes de Informaci\'on, EPN,
Quito,  170525, Ecuador. (e-mail: $\rm jose.vega01@epn.edu.ec$).}


\blfootnote{\noindent F.~J. Lopez-Martinez is with Departamento de Ingenieria de Comunicaciones, Universidad de Malaga - Campus de Excelencia Internacional Andalucia Tech., Malaga 29071, Spain (e-mail: $\rm fjlopezm@ic.uma.es$).}

\vspace{-12.5mm}
\begin{abstract}
Channel modeling is a critical issue when designing or evaluating the performance of reconfigurable intelligent surface (RIS)-assisted communications. Inspired by the promising potential of learning-based methods for characterizing the radio environment, we present a general approach to model the RIS end-to-end equivalent channel using the unsupervised expectation-maximization (EM) learning algorithm. We show that an EM-based approximation through a simple mixture of two Nakagami-$m$ distributions suffices to accurately approximate the equivalent channel, while allowing for the incorporation of crucial aspects into RIS's channel modeling such as beamforming, spatial channel correlation, phase-shift errors, arbitrary fading conditions, and coexistence of direct and RIS channels. Based on the proposed analytical framework, we evaluate the outage probability under different settings of RIS's channel features and confirm the superiority of this approach compared to recent results in the literature.
\end{abstract}


\vspace{-3mm}
\section{Introduction}
Reconfigurable intelligent surfaces (RISs) have been envisioned as a new paradigm to improve the coverage and spectral efficiency of future wireless networks by customizing the propagation radio environment; that is, RIS technology brings intelligence to the physical channel level. An RIS is a metasurface-based device built as a set of low-cost, nearly passive reflecting units that can be configured via an external programmable controller. Depending on the configuration, the RIS is capable of altering the amplitude and/or phase-shift response of the metasurface to modify the behavior of a radio wave that impinges upon it~\cite{Wu}. Owing to its promising features, RISs have been widely investigated in the literature in the context of physical layer security~\cite{plsris,pls2,Weiping}, non-orthogonal multiple access networks~\cite{nomaris}, full-duplex (FD) communication systems~\cite{fdris}, and many others. 

Channel modeling in the context of RIS is a relevant issue, since the achievable performance of RIS-based communications is determined by the distribution of the equivalent channel observed by the receiver. Hence, a good balance between modeling accuracy and mathematical tractability is key for performance analysis purposes. A deep inspection of the RIS-related research in channel modeling reveals that the common assumption of independent and identically distributed (i.i.d.) fading to model the RIS channels is only justified for the sake of mathematical tractability; several relevant examples include somehow idealistic set-ups~\cite{Kudathanthirige,WangmodelinR2,Taoris3} as well as more realistic scenarios that consider hardware impairments and imperfect phase estimation~\cite{Xunoise1,Badiu2020,Qiannoise2}. Very recently, based upon the formulation in~\cite{emilcorr}, a Gamma approximation for the equivalent composite model in RIS-assisted set-ups that explicitly considers the impact of spatial channel correlation in Rayleigh fading was given using the moment-matching (MoM) technique~\cite{Van2020}.

Similarly, most of the aforementioned references are usually restricted to the case of 
Rayleigh fading. Again, such an assumption is taken because of mathematical simplicity rather than based on physically-motivated evidences, specially in line-of-sight (LOS) scenarios. 
The consideration of arbitrary fading conditions for the individual links in the RIS-based set-ups either requires for the use of simple (but not always accurate) approximations based on the Central Limit Theorem (CLT) \cite{Badiu2020}, or come at the price of a rather high mathematical complexity~\cite{Makarfi2020,Trigui2020} using Meijer-G and multivariate Fox-H functions. In practice, the true distribution of the fading links does not exactly belong to a given state-of-the-art model, and the samples of the individual links may not even be available when using existing channel estimation protocols for RIS \cite{Zheng2020}.

Based on the above considerations, and motivated by the potential of RIS to enable practical intelligent radio environments~\cite{Gacanin2020}, we harness the potential of learning methods to unify key factors (e.g., spatial correlation, arbitrary fading of the links, phase-shift noise, and multiple antennas) in the RIS's channel modeling into a single approach without incurring prohibitive complexity. We propose to approximate the exact distribution of the RIS end-to-end channel by a simple mixture of two Nakagami-$m$ distributions, where the fitting parameters are estimated via an unsupervised Expectation-maximization (EM) algorithm. Since EM algorithms are agnostic to the underlying distribution of the sample data, we exemplify how this approach can be applied in two practical scenarios: \emph{(i)} spatially-correlated Rayleigh channels with phase errors, and \emph{(ii)} generalized fading conditions with phase errors. 

In what follows, upper and lower-case bold letters denote matrices and
vectors; $f_{(\cdot)}(\cdot)$ denotes probability density function (PDF); $F_{(\cdot)}(\cdot)$ is the cumulative density function (CDF); $\mathcal{U}[a,b]$ denotes a uniform distribution on $[a,b]$; $\mathcal{C}\mathcal{N}(\cdot ,\cdot )$ is the circularly symmetric Gaussian distribution; $\mathbb{C}$ denotes the complex numbers;  $\mathbb{E}[\cdot]$ is expectation; $\Gamma(\cdot)$ is the gamma function~\cite[Eq.~(6.1.1)]{Abramowitz}; $\Gamma(\cdot,\cdot)$ is the upper incomplete gamma function~\cite[Eq.~(6.5.3)]{Abramowitz}; $\diag\left ( {\bf{x}} \right )$ is a diagonal matrix whose main diagonal is given by ${\bf{x}}$; ${\bf{I}}_{{\rm N}}$ is the identity matrix of size $N\times N$; $\left ( \cdot \right )^{\rm T}$ is the transpose; $\norm{\cdot}$ is the Euclidean norm of a complex vector; $\left ( \cdot \right )^{\rm H}$ is the  Hermitian transpose; $\rm{mod} \left ( \cdot \right )$ is the modulus operation; $\left \lfloor \cdot   \right \rfloor$ is the floor function, and $\sinc(w)=\sin(\pi w)/(\pi w)$ is the sinc function. 
\vspace{-3.5mm}
\section{System and Channel Models}
We consider an RIS-aided wireless communications setup consisting of a transmitter (S) with $M$ antennas communicating with a single-antenna receiver (D) via an RIS equipped with $N$ nearly passive reconfigurable elements. The receiver complex baseband signal at D is expressed as:

\begin{equation}
\label{eq1}
y=\sqrt{P_{\rm T}} \left ( {\bf{h}}_{2}^{\rm T}\bm{\Phi}{\bf{G}} + {\bf{h}}_{\rm sd}^{\rm T} \right ) {\bf{w}} x+\widetilde{n},
\end{equation}
where $P_{\rm T}$ indicates the transmit power at S, $x$ is the transmitted signal with $\mathbb{E}\{|x|^2\}=1$, $\widetilde{n}\sim \mathcal{C}\mathcal{N}(0,\sigma^2_{\widetilde{w}})$ is the additive white Gaussian noise with $\sigma^2_{\widetilde{w}}$ power, ${\bf{h}}_{\rm sd} \in \mathbb{C}^{M\times 1}$ is the direct channel between S and D, $\bf{w}$ $\in \mathbb{C}^{M\times 1}$ is the active beamforming vector at S, ${\bf{G}}= \left [ \bf{g}_1,\dots, \bf{g}_{M} \right ]^{\rm T} \in \mathbb{C}^{N\times M}$ and ${\bf{h}}_{2}=\left [ h_{21},\dots,  h_{2N} \right ]^{\rm T} \in \mathbb{C}^{N\times 1}$ denote the channel coefficients for the S-to-RIS and RIS-to-D links, respectively. Furthermore, $\bm{\Phi}=\diag\left ( e^{j\phi_1},\dots, e^{j\phi_N}\right )$ is the phase-shift matrix induced by the RIS. Let $\angle g_{qn}$ and $\angle h^q_{{\rm sd}}$ be the phase of $g_{qn}$ and $h^q_{{\rm sd}}$ for $q=\left \{ 1,\dots,M \right \}$ and $n=\left \{ 1,\dots,N \right \}$, respectively. So, the RIS uses the optimally-designed phase shifts for each element $\phi_n=\angle h^q_{{\rm sd}}-\angle g_{qn}-\angle h_{2n}$ to cancel the overall phase-shift, which maximizes the SNR at D. Nevertheless, in practice, the imperfect phase estimation and the limited quantization of phase states at the RIS causes that a residual random phase error $\Theta_n$ persists \cite{emilcorr}, i.e., $\phi_n=\angle h
^q_{{\rm sd}}-\angle g_{qn}-\angle h_{2n}+\Theta_n$. Hence, the equivalent magnitude channel observed by D can be formulated as\footnote{To reach the expression in~\eqref{eq2}, we use the equivalent operation, ${\bf{h}}_2^{\rm T}\bm{\Phi} \equiv {\bf{v}}^{\rm T} \diag\left ( {\bf{h}}_2\right )$, where ${\bf{v}}=\left [ e^{j\phi_1},\dots,e^{j\phi_N} \right ]$.}
\begin{equation}
\label{eq2}
h= \left | \left ( {\bf{v}}^{\rm T} \diag\left ( {\bf{h}}_2\right ) {\bf{G}} + {\bf{h}}_{\rm sd}^{\rm T} \right )\bf{w}_{\rm opt} \right |,
\end{equation}
where ${\bf{v}}=\left [ e^{j\Theta_1},\dots,e^{j\Theta_N} \right ]$, with $\Theta_{n}$ being the phase error terms. Under the maximum ratio transmission (MRT), the optimal transmit beamforming vector at S is given by $\bf{w}_{\rm opt} =  \tfrac{\left ( {\bf{v}}^{\rm T} \diag\left ( {\bf{h}}_2\right ) {\bf{G}} + {\bf{h}}_{\rm sd}^{\rm T} \right )^{\rm H}}{\norm{{\bf{v}}^{\rm T} \diag\left ( {\bf{h}}_2\right ) {\bf{G}} + {\bf{h}}_{\rm sd}^{\rm T}} }$ $ \in \mathbb{C}^{M\times 1} $ \cite{MRT}. With the previous definitions, the instantaneous SNR at D is given by
\begin{equation}\label{eq3}
\gamma_{ {\bf{G}},{\bf{h}}_2,{\bf{h}}_{\rm sd}}=\frac{P_{\rm T}}{\sigma_{\widetilde{w}}^2} {{h}}^2.
\end{equation}
Here, our goal is to provide an approximate statistical model for ${{h}}$
and then obtain the SNR distribution straightforwardly. For benchmarking purposes with existing results in the literature, we will consider two rather general situations: $\emph{(i)}$ spatially correlated Rayleigh fading channels, and $\emph{(ii)}$  generalized i.i.d. fading channels. In all instances, S and D are assumed to be well-separated so that their direct channel is independent from the RIS channels. Next, we detail the specific conditions for both scenarios.
\vspace{-3mm}
\subsection{Correlated Rayleigh fading channels}
Here, we take into account spatial Rayleigh correlation for the fading channels ${\bf{G}}$ and ${\bf{h}}_2$. Based on the assumptions in \cite{emilcorr}, for a rectangular phase-shift array with $N=N_V N_H$ elements, where $N_V$ and $N_H$ denotes the number of elements per row and per column, respectively, and under isotropic
scattering environment, the distributions of the fading channels in \eqref{eq1} are described by:
\begin{align}\label{eq4}
{\bf{h}}_{\rm sd}&\sim \mathcal{C}\mathcal{N}\left (\bm{0}_{\rm M}, \beta_{\rm sd}\vspace{0.8mm} {\bf{I}}_{\rm M}\right ), \hspace{2mm}  {\bf{h}}_2\sim \mathcal{C}\mathcal{N}\left ( \bm{0}_{\rm N},A \beta_2 {\bf{R}} \vspace{0.8mm} {\bf{I}}_{\rm N}\right )  \nonumber \\ {\bf{g}}_{\rm q}&\sim \mathcal{C}\mathcal{N}\left ( \bm{0}_{\rm N},A \beta_1 {\bf{R}} \vspace{0.8mm} {\bf{I}}_{\rm N}\right ) \ \text{for} \ q=\left \{ 1,\dots,M \right \},   
\end{align}
where $\beta_1$, $\beta_2$, and $\beta_{\rm sd}$ encompass the average attenuation due to the path losses for the S-RIS, RIS-D, and S-D links, respectively. Also, $A=d_Hd_V$ is
the area of a single RIS element, where $d_V$ is
the vertical height and $d_H$ is the horizontal width, and ${\bf{R}} \in \mathbb{C}^{N\times N}$ denotes the spatial correlation matrix for the RIS. The  $(a,b)$-th element of ${\bf{R}}$ is given by \begin{equation}\label{eq5}
r_{a,b}= \sinc\left ( 2\left \| {\bf{u}}_a-{\bf{u}}_b \right \| /\lambda \right )     \hspace{2mm}  a,b=1,\dots,N
\end{equation}
where ${\bf{u}}_\zeta =\left [ 0, \rm{mod}\left ( \zeta -1,N_H\right )d_H, \left \lfloor \left (\zeta -1 \right )/N_H \right \rfloor d_V \right ]^{T}$, $\zeta  \in \left \{ a,b \right \}$, and $\lambda$ is the wavelength of a plane wave. 
\vspace{-2mm}
\subsection{Generalized i.i.d. fading channels}
We now analyze the case on which the all channel coefficients at the RIS are arbitrarily distributed. For the sake of generality, we consider that the fading channel coefficients are built as a superposition of an arbitrary number $L$ of dominant specular waves plus an additional diffuse components \cite{Durgin2000}, as:
\begin{align}
\label{eq4}  
{h}^q_{\rm sd} =&\sqrt{\beta_{\rm sd}}\left (  \sum_{l=1}^{L} V^{(l,q)}_{\rm sd}e^{j\theta^{(l,q)}_{\rm sd}}+Z^{(q)}_{\rm sd}\right ) \\ {{h}}_{2n} =& \sqrt{\beta_{2}}\left ( \sum_{l=1}^{L} V^{(l,n)}_{2}e^{j\theta_{2}^{(l,n)}}+Z^{(n)}_{2}\right )   \ \\  {{g}}_{qn} =&\sqrt{A\beta_{1}}\left ( \sum_{l=1}^{L} V_1^{(l,q,n)}e^{j\theta_1^{(l,q,n)}}+Z_1^{(q,n)}\right ) 
\end{align}
with $q=\left \{ 1,\dots,M \right \}$ and $n=\left \{ 1,\dots,N \right \}$, and where $V_{(\cdot)}^{(l,\cdot)}$ denote the constant amplitude of an $l$-th specular component, $\theta_{(\cdot)}^{(l,\cdot)} \sim \mathcal{U}[0, 2\pi] $, and $Z_{(\cdot)}^{(\cdot)}$ are Rayleigh distributed with $\mathbb{E}\{|Z|^2\}=2\sigma^2=\Omega_0$ denoting the diffuse received signal components. This formulation includes important ray-based fading models such as Rayleigh, Rician, and two-wave with diffuse power (TWDP) as special cases for $L = 0,1,2$, respectively.

\section{EM-based RIS Channel Modeling}
In this section, we describe how the EM learning algorithm can be used to model the equivalent RIS channel. Even though this general approach can be used for any target distribution, we exemplify that a simple mixture of two Nakagami-$m$ distributions\footnote{Experience shows that, in general, the higher the number $\texttt{Mix}$ of the Nakagami-$m$ distributions being mixed, the better the fit of the approximate solution concerning the true distribution. On the other hand, beyond a certain value of $\texttt{Mix}$, $i)$ EM becomes computationally hard, and $ii)$  mathematical tractability of the approximation may become tedious. Choosing $\texttt{Mix}=2$ is a good rule of thumb that leads to a trade-off among accuracy, computational cost, and mathematical complexity.} provides an excellent performance.
\subsection{Proposed Approximation}
Let us consider a training set vector ${\bf{h}}=\left \{ h_i \right \}_j^t$ consisting of $t$ samples of $h$ in \eqref{eq2} observed by D; we propose to approximate the distribution of $h$ by a mixture
of two Nakagami-$m$ distributions, given by
\vspace{-2mm}
\begin{align}\label{eq7}
f_{h}(r)= &\sum_{i=1}^{2}\omega_{i}\phi_{i}\left (r;m_i,\Omega_i\right )=\sum_{i=1}^{2}\omega_{i}\frac{m_{i}^{m_{i}}r^{2m_i}e^{-\frac{m_{i}r^{2}}{\Omega_{i}}}}{2^{-1}\Gamma (m_{i})\Omega _{i}^{m_{i}} r} ,
\vspace{-3mm}
\end{align}
where $\omega_{i}, \in \left \{ 1,2 \right \}$ under the constraints of $\sum_{i=1}^{2}\omega_{i}=1$ and $0 \leq \omega_{i} \leq1$ denote the mixture weights, $\Omega _{i}$ are the mean powers, and $m_{i}$ are the fading parameters of the weighted PDFs. Moving from the conventional channel modeling approaches to learning-based channel modeling, we adopt a practical fitting technique based on the EM algorithm in order to estimate the mixture parameters in \eqref{eq7} \cite{Aldossari}. EM is an iterative approach that maximizes the mixture model's likelihood function with respect to the weight coefficients using the input unlabeled samples. The EM
algorithm consists of two steps, namely, the expectation (E)-step and the maximization (M)-step. Initially, the parameters in the EM algorithm are randomly chosen for the mixture model. Then, the parameters are updated on each iteration until a convergence criteria is met. E-step calculates the membership coefficients of the $i$th weighted PDF for all data point utilizing the current parameter estimates $\omega_{i}$, $\Omega_{i}$ and $m_i$. The membership values can be computed as \cite{access}
\begin{equation}\label{Responsabilities}
\tau_{ij}^{(k)}=\frac{\omega_{i} \phi_{i}\left ( h_{j};m_i,\Omega_i \right )}{\sum_{l=1}^{2}\omega_{l}\phi_{l}\left ( h_{j};m_i,\Omega_i \right )
	},\hspace{0.05cm} i=1,2,  j=1,2,\dots,t,
\end{equation}
where $k$ denotes the current iteration, $t$ is the size of the sample set in \eqref{eq2}, $h_j, \forall j=1\ldots t$ represent the unlabeled samples, and $i$ is the mixture index. Then, in the M-step, the new parameters (i.e., fading and weight values) are estimated by maximizing the log-likelihood function of each mixture distribution weighted by the membership values. The updated parameters are given by \cite{access}:
 \begin{align}\label{eq9}
 \Omega_{i}^{(k+1)}=&  \frac{\sum_{j=1}^{t}\tau_{ij}^{(k)} h_{j}^2}{\sum_{j=1}^{t}\tau_{ij}^{(k)}} \quad  m_{i}^{(k+1)} =   \frac{1+\sqrt{1+\frac{4\Delta^k_i}{3  }}}{4\Delta^k_i}  \nonumber \\
 \omega_{i}^{(k+1)}=&\frac{\sum_{j=1}^{t}\tau_{ij}^{(k)}}{t}, \Delta^{k}_i=\frac{\sum_{j=1}^{t}\tau_{ij}^{(k)}  \left [ \log(\Omega_i)-\log(x_j^2)\right ]}{\sum_{j=1}^{t}\tau_{ij}^{(k)}}
\end{align}
Algorithm~\ref{algoritmo} shows the Nakagami-$m$ Mixture Model based on the EM approach. Here, the mixture weights are randomly chosen in the range of $[0 \hspace{1mm} 1]$, and the initial values for the fading parameters are computed via the conventional Maximum Likelihood Estimation (MLE). The relative tolerance method is used as stopping criterion. The algorithm stops when a relative tolerance between two parameters (i.e., the old and the new values) is lower than a given threshold, which is experimentally set to $1\times10^{-3}$. 

\begin{remark}\label{remark1}
The PDF approximation in \eqref{eq7} is general and new, and it can be easily used to model any existing fading channel assumed in \eqref{eq2}. More importantly, it can also be used with raw measurement data, without the explicit knowledge of the actual wireless propagation characteristics. Also, the obtained framework is, to the best of our knowledge, the first one in the literature that can be used to characterize the RIS's composite channels as correlated/independent over conventional/generalized fading paths in the presence of phase errors without incurring in additional mathematical complexity.
\end{remark}

\begin{algorithm}[t]\label{algoritmo}
\begin{footnotesize}
\SetAlgoLined
 \SetKwInOut{input}{f}{\textbf{Input:}  ${\bf{h}}\leftarrow $ training set, $\epsilon \leftarrow 1\times10^{-3} $, and initialization of $\omega_i$, $\Omega_{i}$, and  $m_i$ via MLE,\enspace for  $i=1,2$}\;
  \SetKwInOut{input}{f}{\textbf{Output:} $\omega_i$, $\Omega_{i}$, and $m_i$,\enspace for  $i=1,2$}\;
  $k=1$,\enspace  $t=\text{length}({\bf{h}})$\;
 \While{$\Lambda \Omega_i \hspace{2mm} \&\& \hspace{2mm} \Lambda m_i < \epsilon $}{
  $\textbf{ \text{E step:}}$ \\
   \For{ $j=1$; $j$ $<$ $t$; $j$++ }{
   	$\tau_{ij}^{k}=\frac{\omega_i^{k}\phi_{i}\left (h_j;m_i^{k},\Omega_i^{k} \right )}{\sum_{l=1}^{2}\omega_{l}\phi_{l}\left ( h_{j};m_i^{k},\Omega_i^{k} \right )}, \enspace \textit{for}   \enspace  i=1,2$\;
}
 $\textbf{ \text{M step:}}$ \\
 	$\Delta_i^k=\frac{\sum_{l=1}^{t}\tau_{il}^{k} \left (  \log(\Omega_i^k)-\log(h_l^2)\right )}{\sum_{l=1}^{t}\tau_{1l}^{k}}
		$\;
	$\Omega_i^{k+1}=\frac{\sum_{l=1}^{t}\tau_{il}^{k}\times h_l^2}{\sum_{l=1}^{t}\tau_{il}^{k}}$ \ $m_{i}^{k+1}=\frac{1+\sqrt{1+\frac{4\Delta_i^k}{3  }}}{4\Delta_i^k}
		$ \;
	$\omega_i^{k+1}=\sum_{l=1}^{t}\tau_{il}^{k}/t, \enspace \textit{for}   \enspace  i=1,2$\;
   $k=k+1$\;}
 $\textbf{ \text{Stop Criterion Definition:}}$ \\
  $\Lambda m_i =\mid (m_i^{(k+1)}- m_i^{(k)})/m_i^k\mid$ \ $\Lambda \Omega_i =\mid (\Omega_i^{(k+1)}- \Omega_i^{(k)})/\Omega_i^k\mid$\;
 \caption{EM procedural algorithm to estimate $\omega_i$, $\Omega_i$, and $m_i$ of Nakagami-$m$ Finite Mixture Model}
  \end{footnotesize}
\end{algorithm}
\vspace{-4.5mm}
\subsection{Outage Probability Performance}
We consider the outage probability (OP) as the benchmarking metric to evaluate the system performance. Hence, the OP expression for the network is given by the following proposition.
\begin{proposition}\label{Propo1}
The approximate OP expression of the proposed system is given by
\end{proposition}
\vspace{-9mm}
\begin{align}\label{eq11}
   \text{OP}= 1-\sum_{i=1}^{2}\frac{\Gamma\left (m_i,\frac{m_i\left ( 2^{R_{\rm th}}-1 \right )}{\Omega_i P_{\rm T}/\sigma_{\widetilde{w}}^2}  \right )}{\Gamma\left ( m_i \right )},
\end{align}
where $R_{\rm th}$  [b/s/Hz]  is the target rate, and the parameters $\left ( m_i,\Omega_i,\omega_i \right )$ for the  mixture model are estimated by using \eqref{eq9}.
\vspace{-4mm}
\begin{proof}
The OP is defined as the probability that information rate is less than the required threshold information rate ($R_{\rm th}$). Therefore the OP of the system can be formulated as
\begin{align}\label{eq12}
 \text{OP}&=\Pr\left \{ \log_2\left ( 1+ \tfrac{P_{\rm T}}{\sigma_{\widetilde{w}}^2} {h}^2 \right )< R_{\rm th}  \right \}   =F_{{h}}\left ( \sqrt{\tfrac{2^{R_{\rm th}}-1}{P_{\rm T}/\sigma_{\widetilde{w}}^2}} \right ) .
\end{align}
From \eqref{eq12}, the OP in \eqref{eq11} can be obtained directly from the cumulative distribution function (CDF) of \eqref{eq7}.
\end{proof}
\vspace{-3mm}
\section{Numerical results and discussions} \label{sect:numericals}

We now evaluate the effect of correlated/i.i.d channels under traditional/generalized fading with phase errors on the performance in the investigated scenario, as well as the goodness of the proposed approximation for the equivalent channel in RIS-aided system. For all plots, we consider a RIS network geometry as in \cite{emilcorr}, where the fixed system parameters are setting as $P_{\rm T}/\sigma_{\widetilde{w}}^2=124$ dB, which corresponds to transmitting 30 dBm over 10 MHz of bandwidth with 10 dB noise figure, and a carrier frequency of 3 GHz, so the size of a single RIS element will be $\lambda=0.1$ mts. The phase errors $\Theta_n$ are modeled as zero-mean Von Mises RVs with concentration parameter $\kappa$, which captures the accuracy of the phase estimation at the RIS elements (i.e., a smaller $\kappa$ means a larger phase error). For the sake of comparison, the approaches in \cite{Van2020} and \cite{Badiu2020} for modeling correlated Rayleigh and i.i.d. generalized RIS channels, respectively, are included as a reference in the OP analysis. For informative purposes, to estimate the mixture model parameters in \eqref{eq7} with the EM algorithm's aid, $t=10^5$ realizations are generated for the training set in \eqref{eq2} for all instances. Monte Carlo (MC) simulations for the true channels are provided to validate the accuracy of the proposed framework.

\begin{figure*}[ht!]
\vspace{-5.5mm}
    \centering
\psfrag{A}[Bc][Bc][0.5]{$\mathrm{\textit{N}=36}$}
\psfrag{B}[Bc][Bc][0.5]{$\mathrm{\textit{N}=144}$}
    \subfigure[]{\includegraphics[width=0.32\textwidth]{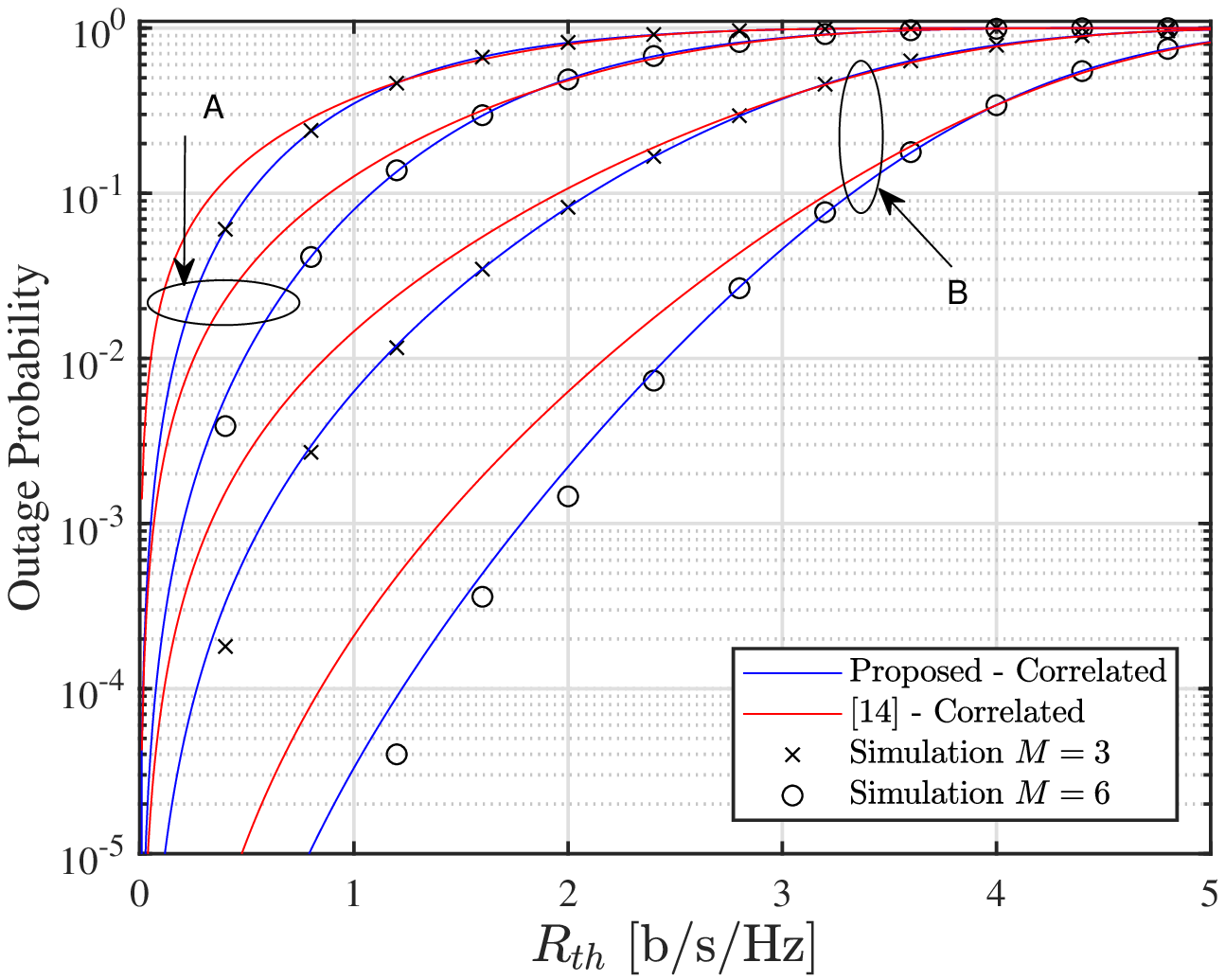}} 
\psfrag{A}[Bc][Bc][0.5]{$\mathrm{\textit{N}=36}$}
\psfrag{B}[Bc][Bc][0.5]{$\mathrm{\textit{N}=100}$}
\psfrag{C}[Bc][Bc][0.5]{$\mathrm{\textit{N}=256}$}
    \subfigure[]{\includegraphics[width=0.32\textwidth]{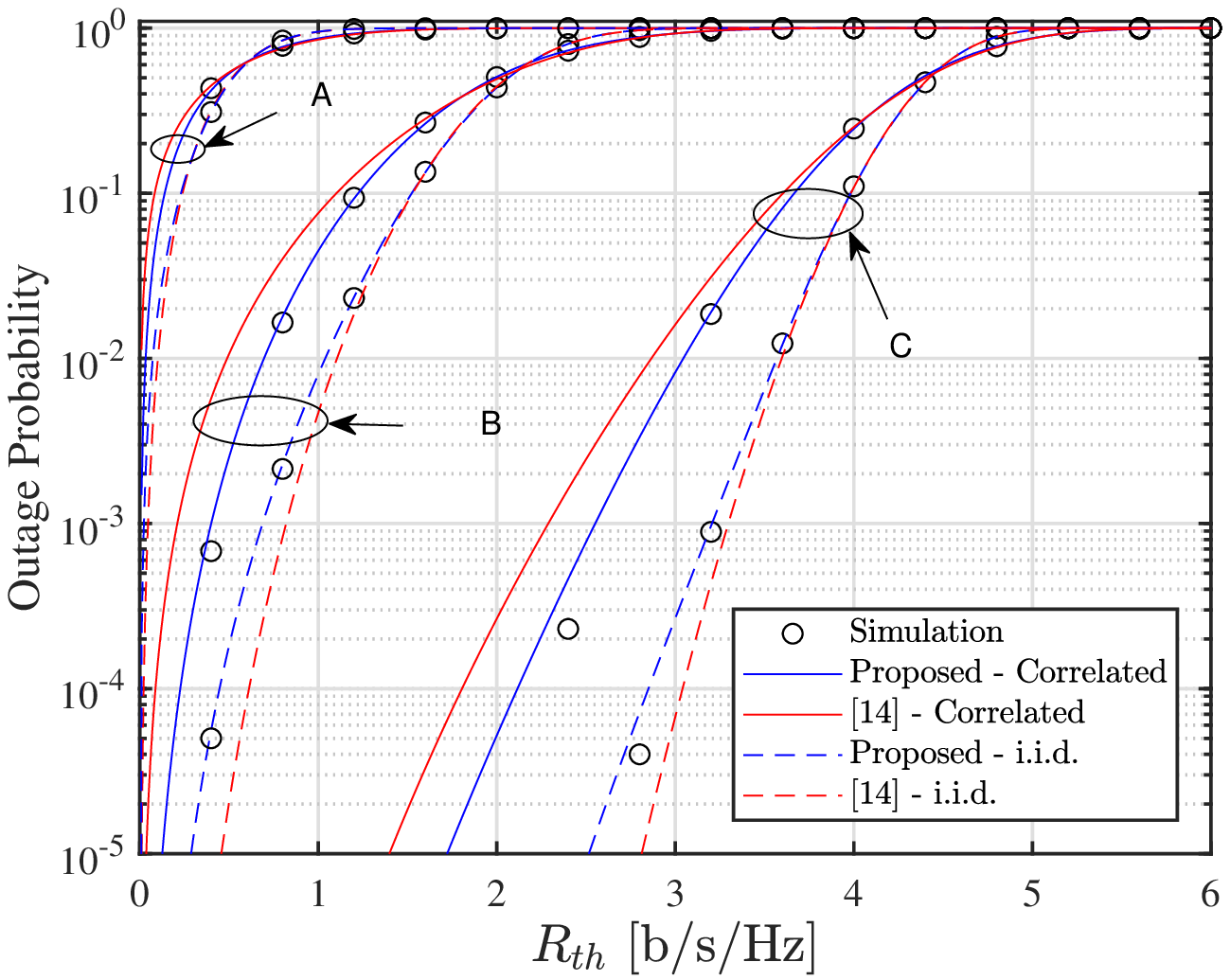}}
\psfrag{A}[Bc][Bc][0.5]{$\mathrm{\lambda/12}$}
\psfrag{B}[Bc][Bc][0.5]{$\mathrm{\lambda/8}$}
\psfrag{C}[Bc][Bc][0.5]{$\mathrm{\lambda/4}$}
\psfrag{D}[Bc][Bc][0.5]{$\mathrm{i.i.d. \hspace{1mm}}\mathrm{fading}$}
    \subfigure[]{\includegraphics[width=0.32\textwidth]{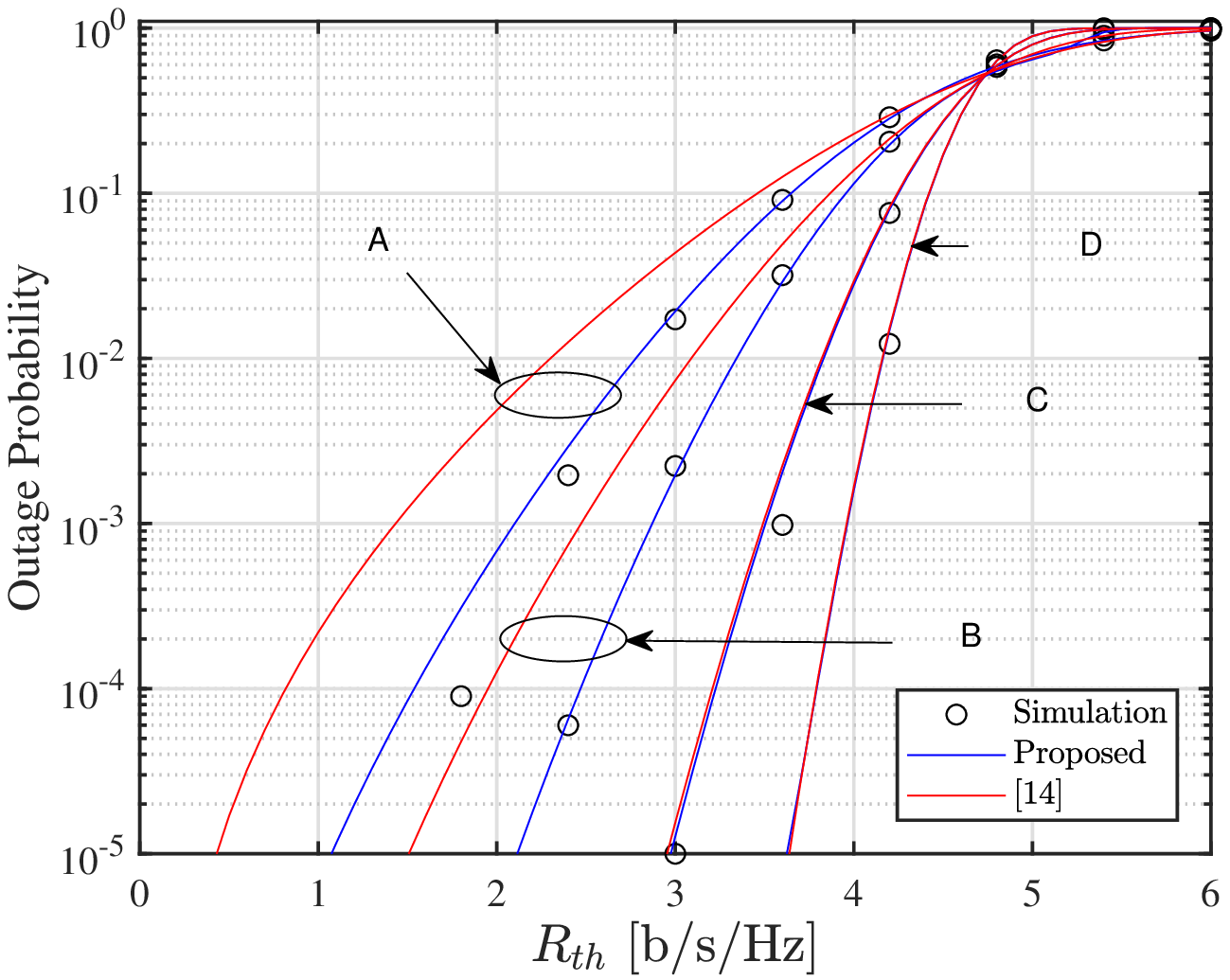}}
    \caption{OP vs. $R_{\rm th}$ by assuming correlated Rayleigh channels in different scenarios: (a) no direct link exists between S and D and varying $N$ and $M$; (b) only the indirect link channel for different $N$ values and $M=1$; (c) the joint presence of direct (S-D) and correlated RIS channels for different correlation matrix distances and $M=1$.}
    \label{fig:foobar}
    \vspace{-4mm}
\end{figure*}

Figs.~1a-1c illustrate the impact of assuming correlated Rayleigh fading in the RIS's composite channel on the OP performance. In these figures, for the indirect channels, we consider $A\beta_1=A \beta_2=-75$ dB. Also, for Figs. 1a-1b, the spatial correlation matrices are formulated assuming $d_H=d_V=\lambda/8$, and $\kappa=1$, which is linked to one of the practical limitations of state-of-the-art implementations arising from hardware impairments at the RIS. In Fig.~1a, we present the OP vs. $R_{\rm th}$ by varying the number of both $N$ (RIS elements) and $M$ (transmit antennas) when the direct
channel is blocked. From all traces, we see that the source's antenna configuration contributes to improving the OP performance as $M$ increases, regardless of the number of RIS elements. Fig.~1b shows the OP vs. $R_{\rm th}$ for different values of $N$ without the existence of the direct link for $M=1$. The i.i.d. Rayleigh case is also included as a reference. From all the curves, it is evident that channel correlation significantly affects the OP compared to the i.i.d. Rayleigh case. In fact, the performance gap between correlated and independent channels becomes more consistent as $N$ increases. 
In Figs. 1a-1b, notice that the differences between MC simulations and the proposed EM-based approximation are almost imperceptible. Conversely, the MoM approach in~\cite{Van2020}, although reasonably good, is notably outperformed by our approximation. Fig.~1c depicts the OP vs. $R_{\rm th}$ for $M=1$ in the presence of both direct and indirect channels. Here, we explore the effect of varying the size of the RIS element on the OP performance. Hence, we use different correlation matrix distances, i.e., $d_H=d_V \in \left \{\lambda/4, \lambda/8, \lambda/12 \right \}$. Also, the other system parameters are set to: $\beta_{\rm sd}=-130$ dB, and $\kappa=3$. As in the previous figure, the case of i.i.d. Rayleigh fading is also reported as a reference. From all traces, it can be observed that decreasing the size of the RIS (i.e., reducing the inter-element distance) leads to a significant loss in the OP performance when dealing with correlated channels; in contrast, the best OP performance is achieved with the unrealistic i.i.d. Rayleigh case. Again, the proposed approach presents a better fit than the approach in \cite{Van2020} to the MC simulations.  

To demonstrate the generality of the proposed EM-based approach, we now study how the consideration of assuming i.i.d. generalized fading channels in a RIS-aided communication impacts the OP behavior. Figs.~2a and~2b show the OP vs. $R_{\rm th}$ by varying the number of elements at the RIS without/with the existence of the direct link, respectively. For the sake of simplicity, we define a power ratio parameter similar to the well-known Rician $K$ parameter, i.e., $K_{L}\buildrel \Delta \over = \frac{\Omega_{L}}{\Omega_0}$, with $\Omega_{L}=\sum_{l=1}^{L}V_{l}^2$ being the total average power of the specular components. Likewise, the amplitudes of the successive rays are expressed in terms of the amplitude of the first dominant component, as in~\cite{nwdppls}, i.e., $V_{l }=\alpha V_{1}$ for $l=\left \{2, \dots, L\right \}$, with $0<\alpha<1$. Considering this, in Fig.~2a, for the links between S and the RIS, and between the RIS and D, we consider $L=2$, $K=2$ dB, $V_1=1$, $\alpha=0.5$, and $\Omega_0=1$. Also, $\beta_1=\beta_2=-55$ dB, and $d_H=d_V=\lambda/2$, and $\kappa=1$. In Fig.~2b, all the previous configurations in Fig.~2a for the indirect channels are kept. Regarding the direct path between S and D, we assume Rician fading, i.e., $L=1$ with $K=5$ dB, and $\beta_{\rm sd}=-135$ dB. Comparing Figs.~2a and~2b, we see a remarkable performance gain due to the existence of a direct link. For instance, when $N=196$ and $R_{\rm th}=0.4$, the OP is $1.3\times10^{-4}$ with the direct link and $3.9\times10^{-5}$ without the direct path. Finally, for all scenarios under study, our approximations work well regardless of $N$, while the approach in \cite{Badiu2020} (vide Fig.~2a) is slightly degraded for low OP values. Finally, Table~\ref{table1} shows the normalized mean square error (NMSE) evaluated using the \texttt{NMSE} built-in function in MATLAB, as a figure of metric to quantify the goodness-of-fit among the MC simulations, the proposed EM method, and the approaches in~\cite{Badiu2020,Van2020} of the curves considered in Figs. 1b and 2a. Results in Table~\ref{table1} support the observation that our approach provides better performance in the left tail (asymptotic region) of the OP plots than its counterparts, regardless of the value of $N$. This difference in accuracy becomes more evident in the illustrative examples shown in the other figures. 
\vspace{-3mm}
\section{Conclusions}
The potential of learning-based methods for modeling the distribution of RIS end-to-end channels has been explored. Specifically, the use of unsupervised EM techniques facilities incorporating key channel aspects such as spatial correlation, presence of phase noise at the RIS, the existence of both the direct and indirect paths, and assumptions of conventional/generalized fading. The potential of using these techniques with channel measurement data opens the possibility to naturally incorporating learning-based techniques in intelligent radio environments.

\begin{figure}[t]
\vspace{-2.5mm}
\centering
\psfrag{A}[Bc][Bc][0.5]{$\mathrm{\textit{N}=49}$}
\psfrag{B}[Bc][Bc][0.5]{$\mathrm{\textit{N}=100}$}
\psfrag{C}[Bc][Bc][0.5]{$\mathrm{\textit{N}=196}$}
\subfigure[no direct link exists between S and D]{\includegraphics[width=65mm]{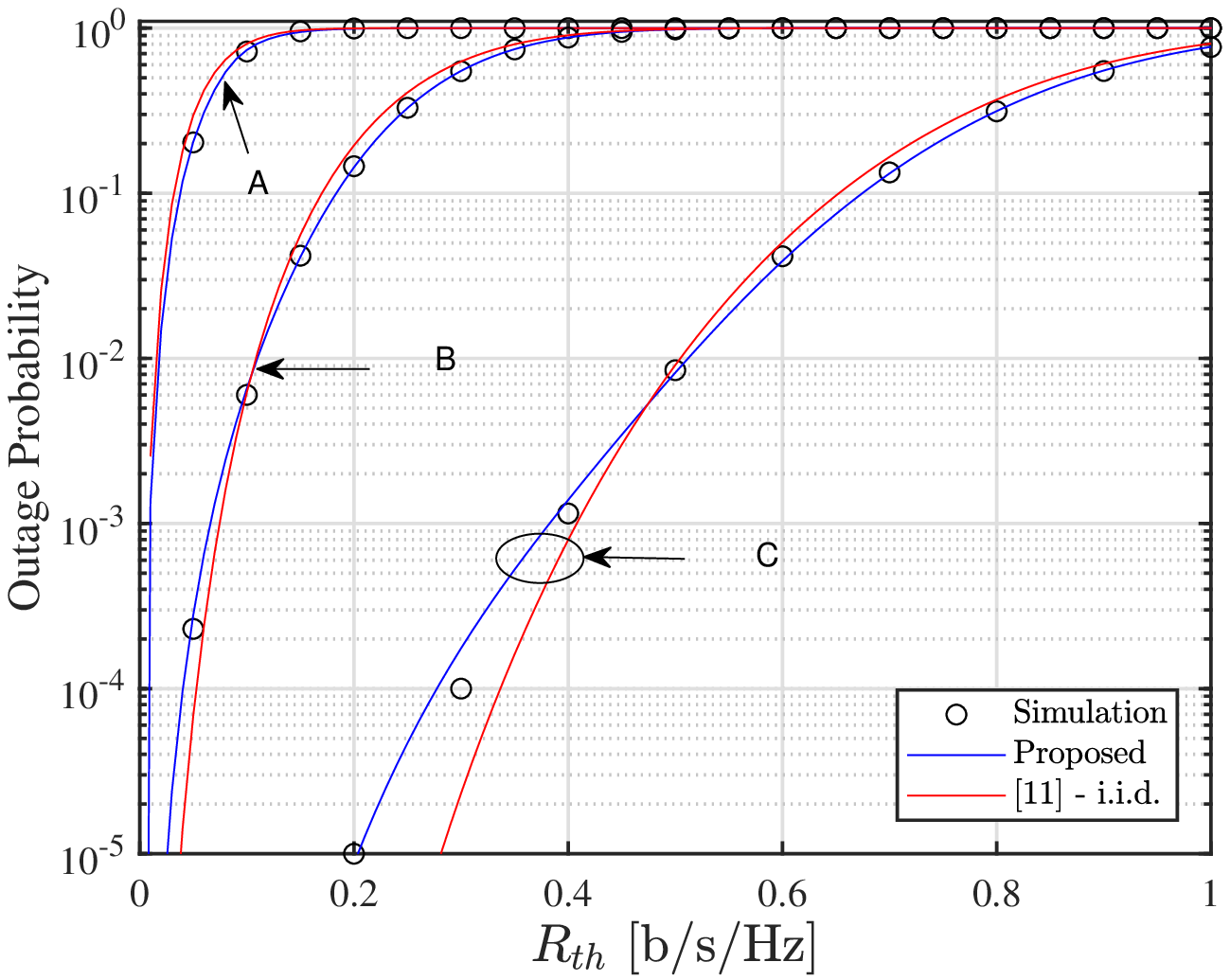}}
\psfrag{A}[Bc][Bc][0.5]{$\mathrm{\textit{N}=49}$}
\psfrag{B}[Bc][Bc][0.5]{$\mathrm{\textit{N}=100}$}
\psfrag{C}[Bc][Bc][0.5]{$\mathrm{\textit{N}=196}$}
\subfigure[both direct and indirect channels are present]{\includegraphics[width=65mm]{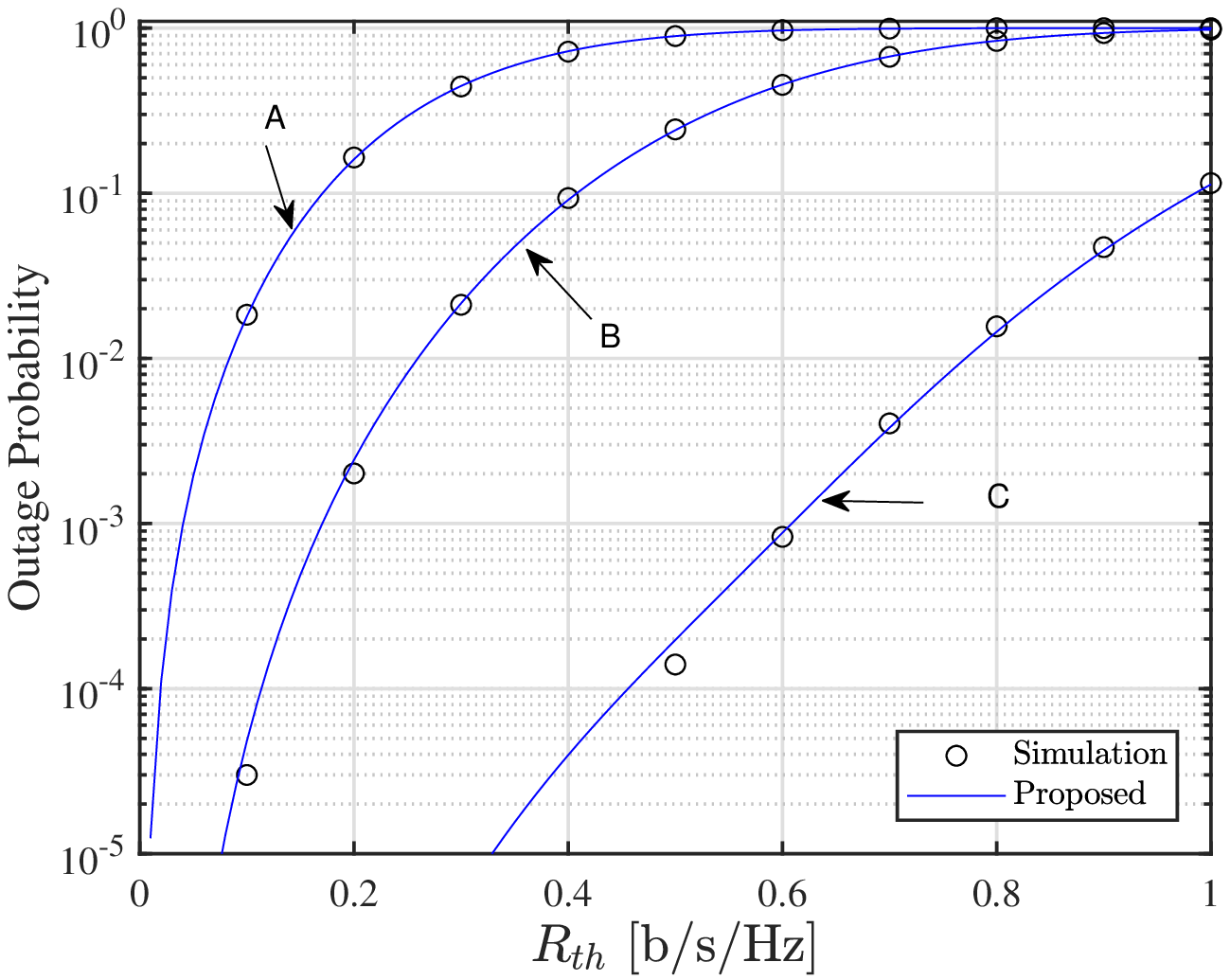}}
\caption{OP vs. $R_{\rm th}$ for different $N$ values and generalized NWDP channels.}
\vspace{-3.5mm}
\label{fig:lego}
\end{figure}

\begin{table}[t!]
    	\caption{OP fitting results. \\NMSE ranges from $-\infty$ (bad fit) to 1 (perfect fit).} 
    \tiny
   \centering
	\begin{tabular}{ccccc}
		\toprule
		\multicolumn{1}{c}{\multirow{2}{*}{}} & \multicolumn{4}{c}{\textbf{\hspace{18mm} NMSE}}\\
		\cmidrule(lr){3-5}
		\textbf{Fig.~\#} &      \textbf{$N$}  &  \textbf{Proposed OP} &  \textbf{OP~\cite{Van2020}-Correlated} & \textbf{OP~\cite{Badiu2020}-i.i.d.} \\
		\cmidrule(lr){1-5}
		\multicolumn{1}{c}{}& 36  & 0.99    &0.96 &-\\
		\cmidrule(lr){2-5}
		\multirow{1}{*}{\textbf{1b}}  &100 & 0.99 & 0.91 &-\\
		\cmidrule(lr){2-5}
		\multirow{1}{*}{\textbf{}}  & 256 & 0.99 &0.86 & - \\
	   \cmidrule(lr){1-5}
		\multicolumn{1}{c}{\textbf{\textcolor{blue}{}}}& 49     &  0.99   & -&0.97 \\
     	\cmidrule(lr){2-5}
     	\multicolumn{1}{c}{\textbf{2a}}& 100    &  0.99   & -& 0.94 \\
     	\cmidrule(lr){2-5}
     	\multicolumn{1}{c}{\textbf{\textcolor{blue}{}}}& 196     &  0.99   & -& 0.88 \\
     	\cmidrule(lr){1-5}
	\end{tabular}\label{table1}
	\vspace{-5mm}
\end{table}
\vspace{-3.5mm}
\bibliographystyle{ieeetr}
\bibliography{bibfile}

\end{document}